\documentclass[epj,twocolumn]{webofc}
\usepackage{bm}
\usepackage{graphicx}
\usepackage[varg]{txfonts}  
\woctitle{TRANSVERSITY 2014}
\begin{document}
\title{Azimuthal asymmetries in {\bm{$p^{\uparrow} p\,\to\, {\rm jet}\, \pi\, X$}}}
\author{Cristian Pisano\inst{1}\fnsep\thanks{\email{cristian.pisano@nikhef.nl}}
\and 
 Umberto D'Alesio\inst{2,3}\fnsep\thanks{\email{umberto.dalesio@ca.infn.it}} 
\and
 Francesco Murgia\inst{3}\fnsep\thanks{\email{francesco.murgia@ca.infn.it}}}

\institute{
Nikhef and Department of Physics and Astronomy, VU University Amsterdam, De Boelelaan 1081, NL-1081 HV Amsterdam, The Netherlands      
\and
Dipartimento di Fisica, Universit\`a di Cagliari, Cittadella Universitaria, I-09042 Monserrato (CA), Italy
\and Istituto Nazionale di Fisica Nucleare, Sezione di Cagliari, C.P. 170,
 I-09042 Monserrato (CA), Italy}

\abstract{We study the azimuthal asymmetries for the distributions of leading pions inside a jet produced inclusively in high-energy proton-proton collisions within the framework of the transverse momentum dependent generalized parton 
model. We present results for the RHIC center-of-mass energies $\sqrt{s} = 200$ and $500$ GeV, mainly for forward jet rapidities, in particular for the two 
mechanisms which dominate such asymmetries: the Sivers and the Collins effects. We 
also briefly discuss the case of inclusive jet production and, adopting 
the so-called colour gauge invariant parton model, we propose a phenomenological analysis of the process dependence of the quark Sivers function.}    
\maketitle
\section{Introduction}
\label{intro}

Polarization phenomena in high-energy hadronic reactions have gathered considerable attention in the last few years from both theoretical and experimental communities~\cite{D'Alesio:2007jt,Barone:2010zz}. Especially the huge single spin asymmetries measured in inclusive forward production of pions in proton-proton collisions are extremely interesting observables, since they cannot be explained in the  usual framework  of leading-twist, perturbative QCD, based on collinear factorization theorems. Within the transverse momentum dependent (TMD) generalized 
parton model (GPM), which takes into account spin and intrinsic parton motion 
effects assuming the validity of QCD factorization, these asymmetries are generated by TMD polarized partonic distribution and fragmentation functions (or TMDs, in short). The most relevant ones from the phenomenological point of view are the quark and gluon Sivers distributions~\cite{Sivers:1989cc} and, for transversely polarized quarks, the Boer-Mulders distribution~\cite{Boer:1997nt} and the Collins fragmentation function~\cite{Collins:1992kk}. Similar functions can be defined for linearly polarized gluons as well~\cite{Mulders:2000sh}. 

In this context, the azimuthal asymmetries in the distribution of pions inside 
a jet with a large transverse momentum are quite interesting observables~\cite{D'Alesio:2010am,D'Alesio:2011mc,D'Alesio:2013jka}, and are presently under active investigation at the Relativistic Heavy Ion Collider (RHIC)~\cite{Poljak:2011vu,Fatemi:2012ry}. In fact, by taking suitable moments of these asymmetries, one could discriminate among the effects due to the different TMDs, in close analogy with the semi-inclusive deep inelastic scattering (SIDIS) case. This is not possible 
for inclusive  pion production, where several underlying mechanisms 
(mainly the Sivers and Collins effects) cannot be separated. In principle, quark and gluon originating jets can also be distinguished. Moreover, one can gain information on the size and sign of TMD distributions and fragmentation functions in kinematic domains in which they are still poorly known. Therefore the study of these observables will definitely be useful in clarifying the role played by the Sivers distribution and by the Collins fragmentation function in the single-spin asymmetries observed for single inclusive pion production. 
We notice that, in a similar analysis that focussed mainly on the universality properties of the Collins function~\cite{Yuan:2007nd}, the transverse partonic motion was considered only in the fragmentation process. In principle our approach has a richer structure in the observable azimuthal asymmetries, because intrinsic motion is taken into account in the initial hadrons as well. However TMD factorization has not been proven for the specific reaction under study, but it is taken as a reasonable phenomenological assumption. Hence the validity of this model still needs to be confirmed by further comparison with experiments.

Finally, we present an extension of the GPM, named colour gauge invariant GPM, which takes into account the effects of initial and final state interactions among active partons and parent hadrons. Such interactions could play a fundamental 
role for the nonvanishing of single spin asymmetries. As a main application, we 
will study the process dependence of the Sivers function for quarks 
in both jet-pion and inclusive pion production at RHIC.

\section{The Generalized Parton Model}

We consider the process
\begin{equation}
p (p_A; S) \,+\, p(p_B)\, \to \, {\rm jet}(p_{\rm j})\,
+\pi(p_\pi)\, +\, X\,,
\end{equation}
where the four-momenta of the particles are given within brackets and one of 
the initial protons is in a pure transverse spin state denoted by the 
four-vector $S$, such that $S^2=-1$ and $p_A\cdot S =0$. All the other particles in the reaction are unpolarized. In the center-of-mass frame of the two incoming protons, $s = (p_A+p_B)^2$ is the total energy squared. Furthermore, we assume that the polarized proton moves along the  positive direction of the $\hat{\bm{Z}}_{\rm cm}$ axis. The production plane containing the colliding beams and the observed jet is taken as the $(XZ)_{\rm cm}$ plane, with $(\bm{p}_{\rm j})_{X_{\rm cm}}>0$. 
In this frame $S = (0, \cos\phi_{S},\sin\phi_{S},0) $ and $p_{\rm j} = 
p_{{\rm j}\,T}(\cosh \eta_{\rm j},1,0,\sinh \eta_{\rm j})$,
with $\eta_{\rm j} = -\log[\tan(\theta_{\rm j}/2)]$ being the jet rapidity.

To leading order in perturbative QCD, the reaction proceeds via the hard 
scattering partonic subprocesses $ab\to cd$, where the final parton $c$ 
fragments into the observed hadronic jet. The corresponding single transversely
polarized cross section has been calculated within the GPM approach, using the 
helicity formalism. Further details can be found in 
Ref.~\cite{D'Alesio:2010am}. 
The final expression has the following general structure, 
\begin{eqnarray}
\hspace{-0.85cm}2{\rm d}\sigma(\phi_{S},\phi_\pi^H)\!\!\!\! &\sim &\!\!\!\!\! {\rm d}\sigma_0
 + {\rm d}\Delta\sigma_0\sin\phi_{S}  + {\rm d}\sigma_1\cos\phi_\pi^H \,\nonumber \\
& + & \!\!\!\!\!
{\rm d}\sigma_2\cos2\phi_\pi^H + {\rm d}\Delta\sigma_{1}^{-}\sin(\phi_{S}-\phi_\pi^H) \nonumber \\
& +& \!\!\!\!\!
{\rm d}\Delta\sigma_{1}^{+}\sin(\phi_{S}+\phi_\pi^H) + {\rm d}\Delta\sigma_{2}^{-}\sin(\phi_{S}-2\phi_\pi^H) \nonumber \\
&+& \!\!\!\! \!{\rm d}\Delta\sigma_{2}^{+}\sin(\phi_{S}+2\phi_\pi^H)\,,
\label{d-sig-phi-SA}
\end{eqnarray} 
where $\phi_\pi^H$ is the azimuthal angle of the three-momentum of the pion around the jet axis, measured in the helicity frame of the fragmenting parton $c$~\cite{D'Alesio:2010am}. 

The different angular modulations of the cross section can be singled out by 
defining the azimuthal moments
\begin{equation}
A_N^{W(\phi_{S},\phi_\pi^H)}
=
2\,\frac{\int{\rm d}\phi_{S}{\rm d}\phi_\pi^H\,
W(\phi_{S},\phi_\pi^H)\,{\cal N}(\phi_{S},\phi_\pi^H)}
{\int{\rm d}\phi_{S}{\rm d}\phi_\pi^H\,{\cal D}(\phi_S, \phi_\pi^H)}\,,
\label{gen-mom}
\end{equation} 
where $W(\phi_{S},\phi_\pi^H)$ is one of the circular functions that appear in 
Eq.~(\ref{d-sig-phi-SA}), while the numerator ${\cal N}(\phi_{S},\phi_\pi^H)$
and denominator ${\cal D}(\phi_{S},\phi_\pi^H)$  of the asymmetries are given 
respectively by
\begin{eqnarray}
\hspace{-0.8cm}{\cal N}(\phi_{S}, \phi_\pi^H) \!\!\!&\equiv& \!\!\! {\rm d}\sigma(\phi_{S},\phi_\pi^H)-
{\rm d}\sigma(\phi_{S}+\pi,\phi_\pi^H) \nonumber \\
 & \sim & \!\!\! {\rm d}\Delta\sigma_0\sin\phi_{S} +
{\rm d}\Delta\sigma_{1}^{-}\sin(\phi_{S}-\phi_\pi^H) \nonumber \\
& +& \!\!\! {\rm d}\Delta\sigma_{1}^{+}\sin(\phi_{S}+\phi_\pi^H) +{\rm d}\Delta\sigma_{2}^{-}\sin(\phi_{S}-2\phi_\pi^H) \nonumber \\
& + &\!\!\! {\rm d}\Delta\sigma_{2}^{+}\sin(\phi_{S}+2\phi_\pi^H)\,,
\label{num-asy-gen}
\end{eqnarray}
and
\begin{eqnarray}
\hspace{-0.5cm}{\cal D}(\phi_S, \phi_\pi^H) & \equiv& {\rm d}\sigma(\phi_{S},\phi_\pi^H)+
{\rm d}\sigma(\phi_{S}+\pi,\phi_\pi^H) \nonumber \\ 
 & \equiv & 2{\rm d}\sigma^{\rm unp}(\phi_\pi^H) \nonumber \\
& \sim & 
{\rm d}\sigma_0 + {\rm d}\sigma_1\cos\phi_\pi^H+
{\rm d}\sigma_2\cos2\phi_\pi^H\,.
\label{den-asy-gen}
\end{eqnarray}

In Ref.~\cite{D'Alesio:2010am} we provide estimates for the upper bounds of all 
the azimuthal moments in Eqs.~(\ref{gen-mom}) in the kinematic regions 
currently under investigation at RHIC. In the following we will focus only on 
those asymmetries that are sizeable, {\it i.e.}~those involving the Sivers and the Collins functions. These TMDs are known and their parameterizations have been extracted from 
independent fits to $e^+e^-$ and  SIDIS data.

\begin{figure*}[t]
\centering
\includegraphics[width=6.5cm]{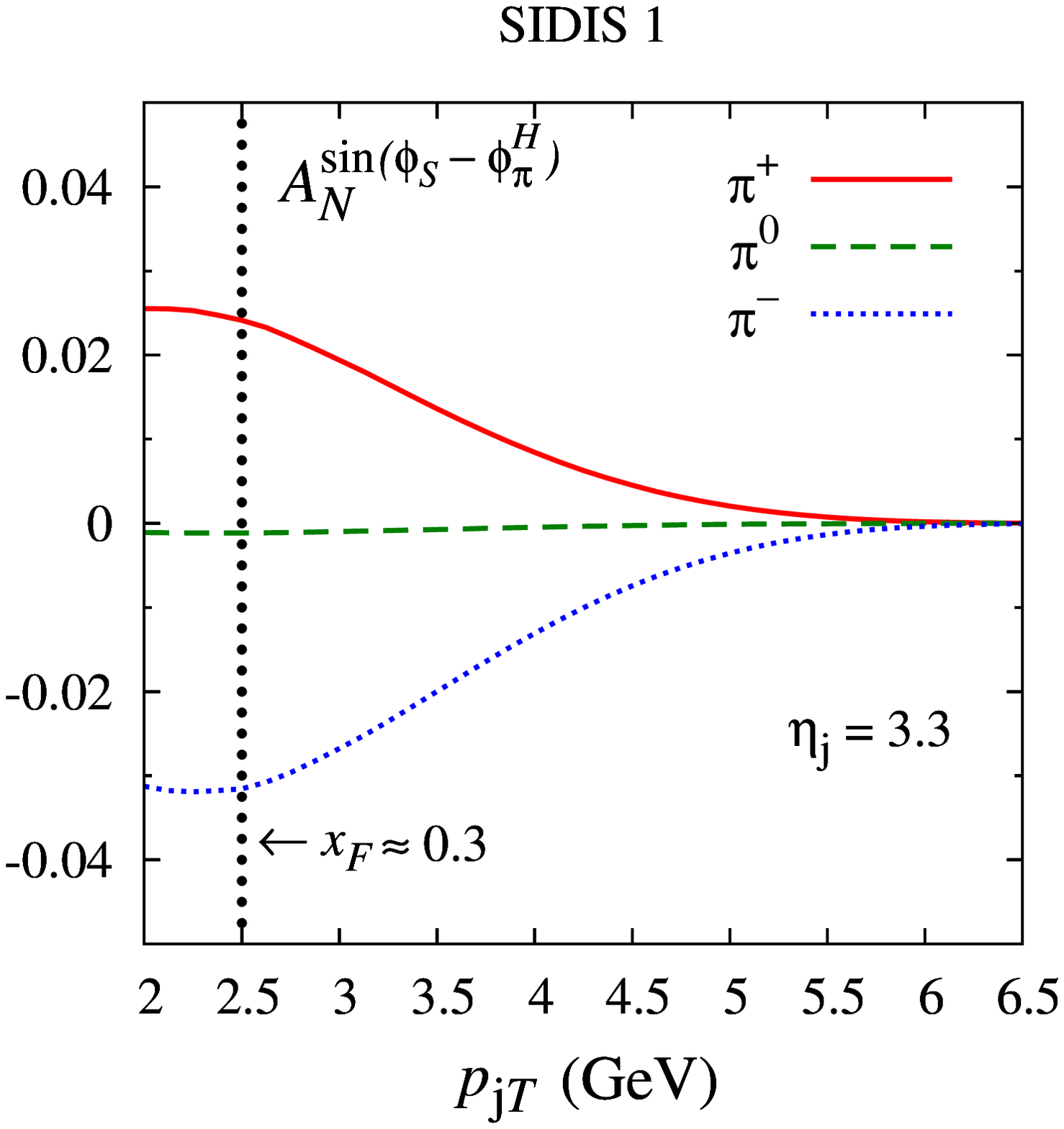}
\includegraphics[width=6.5cm]{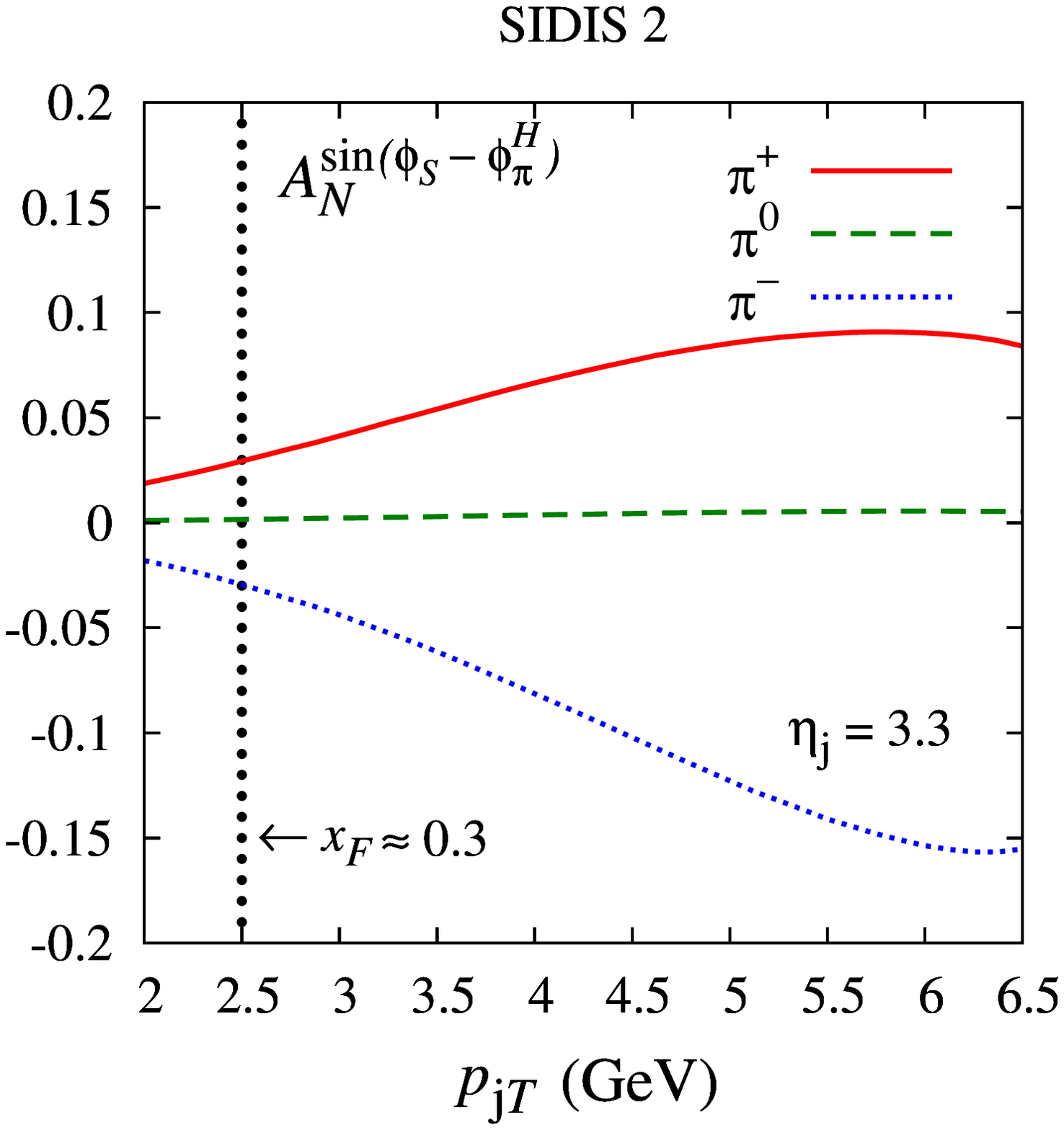}
\caption{The Collins asymmetry $A_N^{\sin(\phi_S-\phi_\pi^H)}$ for the process 
$p^{\uparrow}p\to {\rm jet}\,\pi\,X$ at $\sqrt{s}= 200$ GeV and fixed jet rapidity $\eta_{\rm j}= 3.3$, as a function of the transverse momentum of the jet
$p_{{\rm j} T}$. Results are obtained in the GPM approach, using the sets of
parameterizations SIDIS~1 (left panel) and SIDIS~2 (right panel).}
\label{asy-an-coll-par200}
\end{figure*}

\section{Phenomenology}

In this section the Collins and Sivers asymmetries are evaluated in the 
GPM approach at the RHIC energies $\sqrt s = 200$ GeV and  $\sqrt s = 500$ GeV at forward jet rapidity. Additional phenomenological results can be found 
in Refs.~\cite{D'Alesio:2010am,D'Alesio:2013jka}  

The different TMDs are taken to be universal and are parameterized with a simplified functional dependence on the parton light-cone momentum fraction and on the transverse motion, which are completely factorized. Furthermore, we assume a Gaussian-like flavour-independent shape for the transverse momentum component.
In the following we adopt mainly two different sets of parameterizations, named SIDIS~1~\cite{Anselmino:2005ea,Anselmino:2007fs} and SIDIS~2~\cite{Anselmino:2008sga,Anselmino:2008jk}, described in detail in Ref.~\cite{D'Alesio:2010am}. 
Very recently, updated parameterizations of the transversity distribution
and of the Collins function within the GPM framework have been 
released~\cite{Anselmino:2013vqa}. Since they are anyway qualitatively similar to the ones adopted here, they will not be used in the following.

The hard scale in the process is identified with the jet transverse momentum 
and, since it covers a significant range, the QCD evolution of all TMDs should be taken into account properly. However, 
a formal proof of the TMD factorization for 
this process is still missing and the study of TMD evolution is currently at 
an early stage. Therefore, we tentatively take into account proper evolution
with scale, at leading order in perturbative QCD, only for the collinear parton distribution and fragmentation functions, while keeping fixed the transverse momentum component of all TMDs. 

In all our subsequent predictions, the transverse momentum of the observed pion
with respect to the jet axis (denoted by $\bm{k}_{\perp\pi}$) is integrated out. Moreover, since we are interested in leading particles inside the jet, we integrate over the light-cone momentum fraction of the observed hadron, $z$, in the range $z\geq 0.3$. Different choices can be easily implemented in our numerical calculations, according to the kinematic cuts of interest in specific experiments.

\subsection{The Collins Asymmetries}
\label{sec:coll}

The Collins fragmentation function $H_1^{\perp q}$ contributes to the azimuthal moments $A_N^ {\sin(\phi_S+\phi_\pi^H)}$ and $A_N^ {\sin(\phi_S-\phi_\pi^H)}$ defined in 
Eq.~(\ref{gen-mom}). The first one can schematically be written as
\begin{eqnarray}
\hspace{-0.7cm} A_N^ {\sin(\phi_S+\phi_\pi^H)}\!\!\!\!\!\!&\sim&\!\!\!\!\!\!
\left [ h_{1 T}^{\perp q}(x_a,\bm{k}_{\perp a}^2)\otimes f_1(x_b,\bm{k}_{\perp b}^2) \right .  \nonumber \\
&+& \!\!\!\!\!\left .f_{1 T}^{\perp}(x_a,\bm{k}_{\perp a}^2) \otimes h_1^{\perp q}(x_b,\bm{k}_{\perp b}^2) \right ]
  \otimes H_1^{\perp q}(z,\bm{k}_{\perp \pi}^2)\,.\nonumber \\
\label{eq:Coll1}
\end{eqnarray}
Here, similarly to $z$ and $\bm{k}_{\perp \pi}$ already defined above, we have 
introduced the variables $x_{a,b}$ and $\bm{k}_{\perp a,b}$. These are, respectively, the light-cone momentum fractions and the intrinsic transverse momenta of the incoming partons $a$ and $b$. In the first term on the RHS of Eq.~(\ref{eq:Coll1}), $H_1^{\perp q}$ is convoluted with the unpolarized ($f_1$) and the pretzelosity ($h_{1T}^{\perp q}$) distributions. The second convolution involves the Sivers ($f_{1 T}^{\perp}$) and Boer-Mulders ($h_{1}^{\perp q}$) functions instead. The upper bound of this asymmetry turns out to be negligible \cite{D'Alesio:2010am}.  A similar result holds for  $A_N^ {\sin(\phi_S+2\phi_\pi^H)}$, related to the fragmentation function of linearly polarized gluons. On the other hand, the upper bound of the azimuthal moment 
\begin{equation}
A_N^{\sin(\phi_S-\phi_\pi^H)} \sim h_1^q(x_a,\bm{k}_{\perp a}^2) \otimes
 f_1(x_b,\bm{k}_{\perp b}^2) \otimes H_1^{\perp\, q}(z,\bm{k}_{\perp \pi}^2)\,,
\label{eq:Coll2}
\end{equation}
which is dominated by a convolution of the transversity distribution for 
quarks, $h_1^q$, and the Collins function, is sizeable in the kinematic region
accessible at RHIC~\cite{D'Alesio:2010am}. Analogous conclusion holds for 
its gluonic counterpart $A_N^{\sin(\phi_S-2\phi_\pi^H)}$.

Our estimates for  $A_N^{\sin(\phi_{S}- \phi_\pi^H)}$ are presented in Fig.~\ref{asy-an-coll-par200}, at the hadronic center-of-mass energy $\sqrt{s} = 200$ GeV and  fixed jet rapidity $\eta_{{\rm j}}$= 3.3, as a function of the transverse momentum of the jet, $p_{{\rm j}T}$. These results are obtained by adopting the two parameterizations SIDIS~1 and  SIDIS~2. Recent preliminary data from the STAR 
Collaboration~\cite{Poljak:2011vu} seem to confirm our prediction of an almost vanishing asymmetry for neutral pions. Moreover, we point out that the results for charged pions, obtained with these two different parameterizations, are comparable only in the kinematic region where $x_F=2 p_{{\rm j} L}/\sqrt{s} \le 0.3$ (notice the different scales used in the two panels of Fig.~\ref{asy-an-coll-par200}). This corresponds to the  Bjorken $x$ domain covered by the SIDIS data that have been used to extract the transversity distribution.
Therefore, extrapolations beyond $x_F \approx 0.3$ lead to very different estimates at large $p_{{\rm j} T}$. Consequently, future measurements of the Collins asymmetries for charged pions in this yet unexplored region would shed light on the large $x$ behaviour of the quark transversity distributions.

\begin{figure}[b]
\centering
\includegraphics[width=7cm]{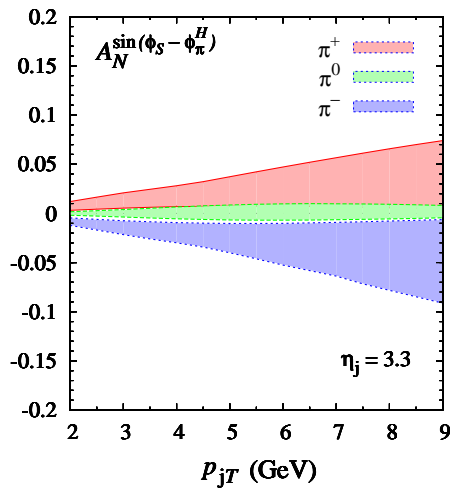}
\caption{Scan bands describing the uncertainty of the Collins asymmetry $A_N^{\sin(\phi_S-\phi_\pi^H)}$ for the process $p^{\uparrow}p\to {\rm jet}\,\pi\,X$ at $\sqrt{s}= 500$ GeV and fixed jet rapidity $\eta_{\rm j}= 3.3$, as a function of the transverse momentum of the jet
$p_{{\rm j} T}$. }
\label{fig:coll-scan}      
\end{figure}

\begin{figure*}[t]
\centering
\includegraphics[width=5.5cm]{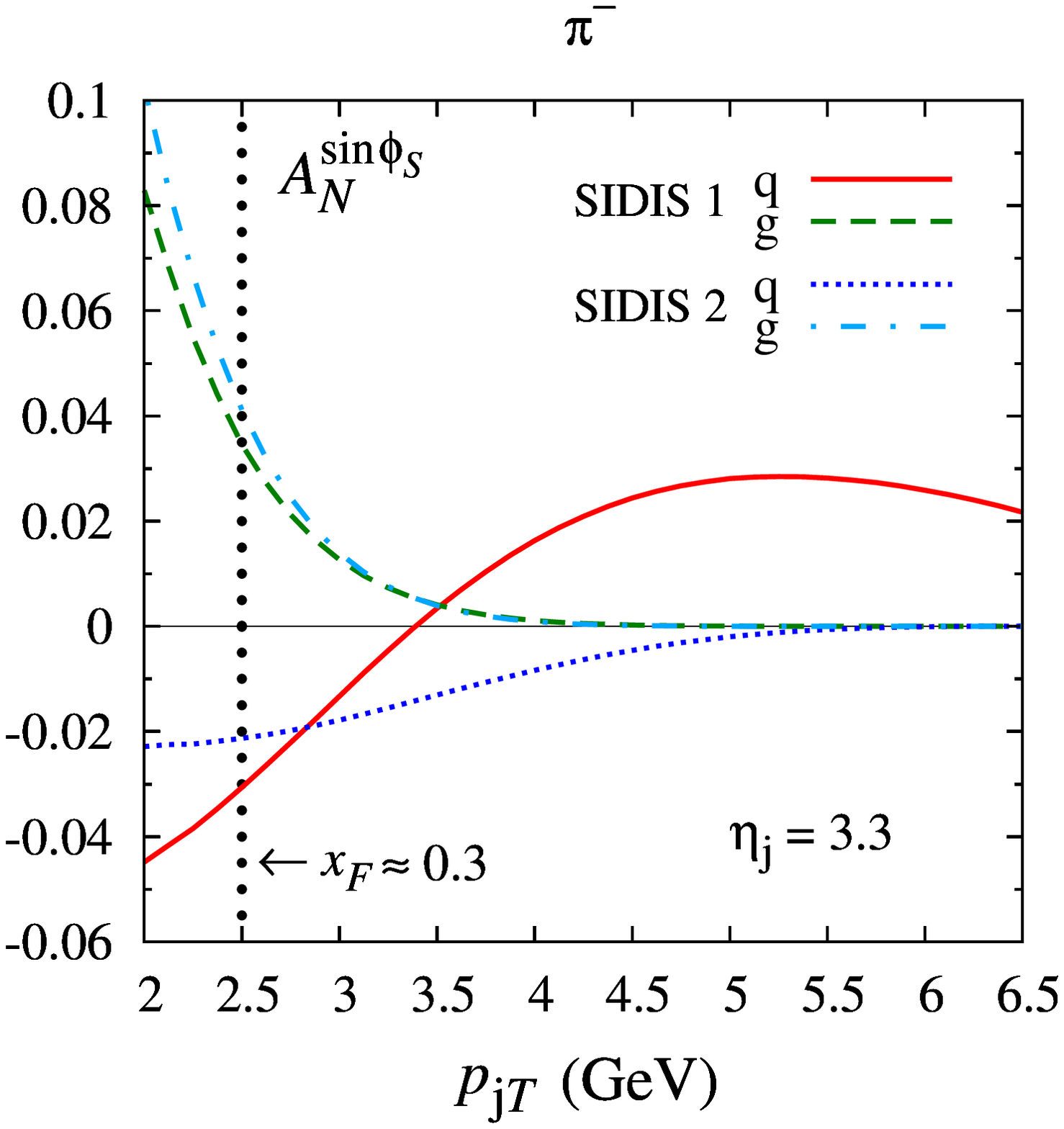}
\includegraphics[width=5.5cm]{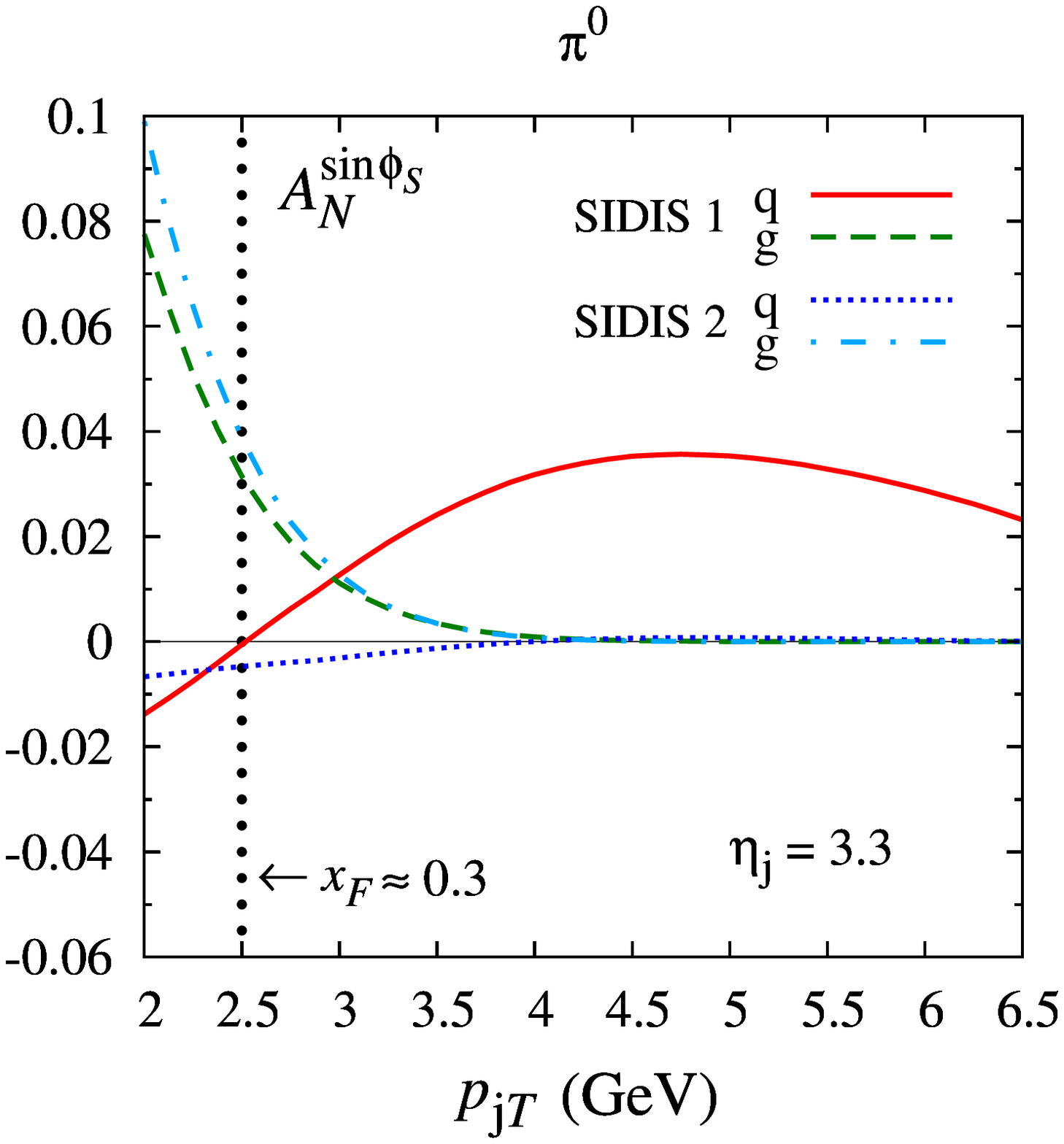}
\includegraphics[width=5.5cm]{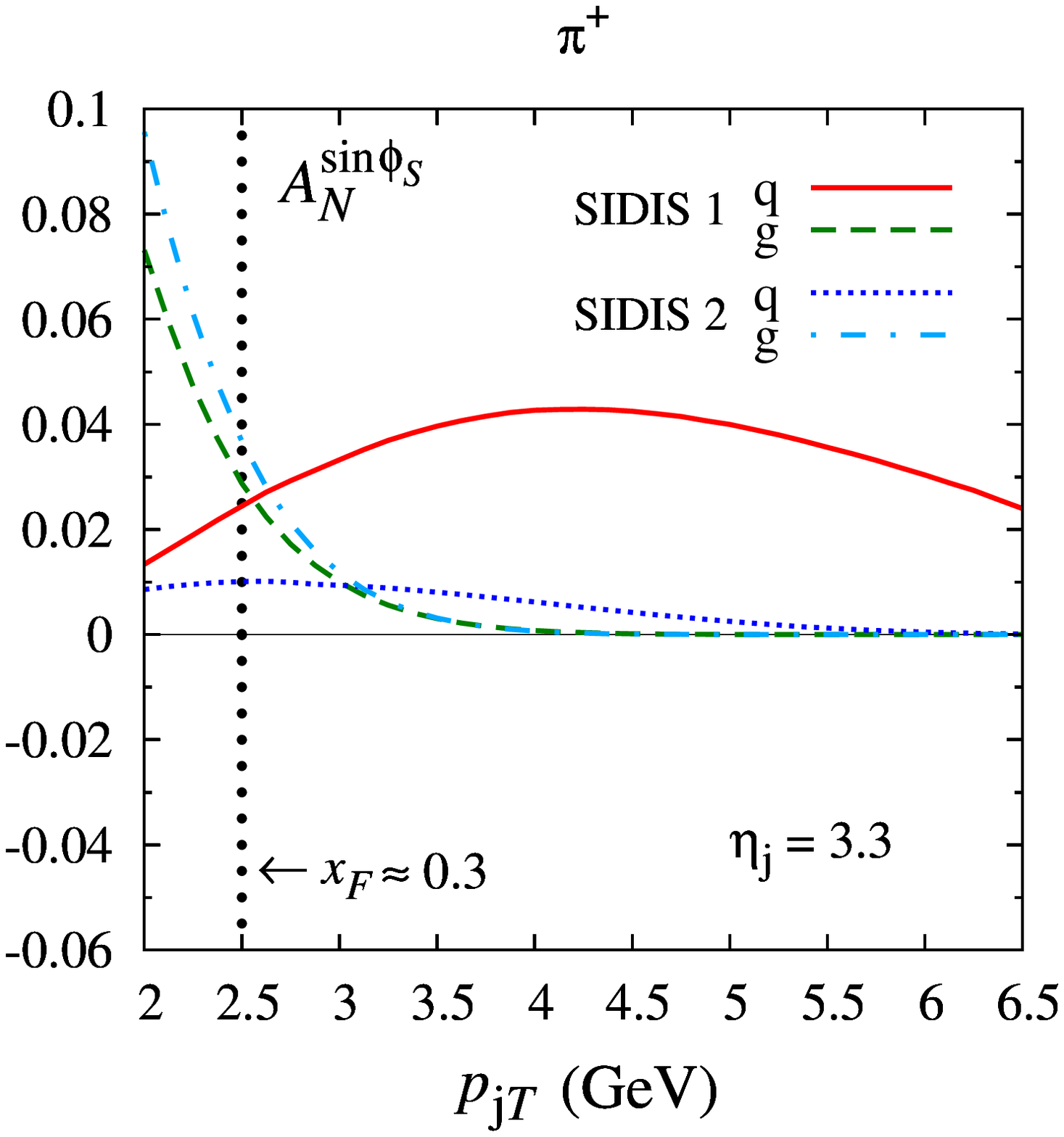}
\caption{The Sivers asymmetry $A_N^{\sin\phi_S}$ for the process 
$p^{\uparrow}p\to {\rm jet}\,\pi\,X$ at $\sqrt{s}= 200$ GeV and fixed jet rapidity $\eta_{\rm j}= 3.3$, as a function of the transverse momentum of the jet
$p_{{\rm j} T}$. Results are obtained in the GPM approach, using the sets of
parameterizations SIDIS~1 and SIDIS~2 for the quark Sivers function. The gluon 
Sivers function is assumed to be positive and to saturate an updated version of
the bound estimated in Ref.~\cite{Anselmino:2006yq}.}
\label{asy-an-siv-par200}
\end{figure*}
Based on the discussion above, we have carried out a complementary study of the uncertainties of our predictions, following the analysis performed in Ref.~\cite{Anselmino:2012rq} within the context of $A_N$ in $p^\uparrow p\to \pi\,X$. We start from a reference fit to {\em updated} SIDIS and $e^+e^-$ data with a total $\chi^2$ denoted by $\chi^2_0$. The resulting parameterizations are therefore slightly different from the SIDIS~1 set, although the same collinear parton distribution and fragmentation functions have been adopted. As a second step, the parameters $\beta_{u,d}$ are fixed within the range $[0,4]$ by discrete steps of $0.5$. These are the parameters that control the large $x$ behaviour of the quark transversity distributions in the factor $(1-x)^{\beta_q}$
of the corresponding  parameterizations~\cite{D'Alesio:2010am}. In this way a total of eighty-one different $\{\beta_u,\beta_d\}$ configurations are obtained. Subsequently, a new fit of the other parameters is performed for each of these $\{\beta_u,\beta_d\}$ pairs, and the corresponding total $\chi^2$ is evaluated. Only those configurations with a 
$\Delta\chi^2=\chi^2-\chi^2_0$ less than a statistically significant reference value \cite{Anselmino:2012rq} are not rejected. It turns out that, in this case,
all eighty-one configurations fulfill the selection criterium. This confirms
our conclusion that presently available SIDIS data do not constrain the large 
$x$ behaviour of the transversity distributions. The final step consists in 
taking the full envelope ({\em scan bands}) of the values of the asymmetry for the process under study, corresponding to the selected configuration sets.

\begin{figure*}[t]
\centering
\includegraphics[width=5.5cm]{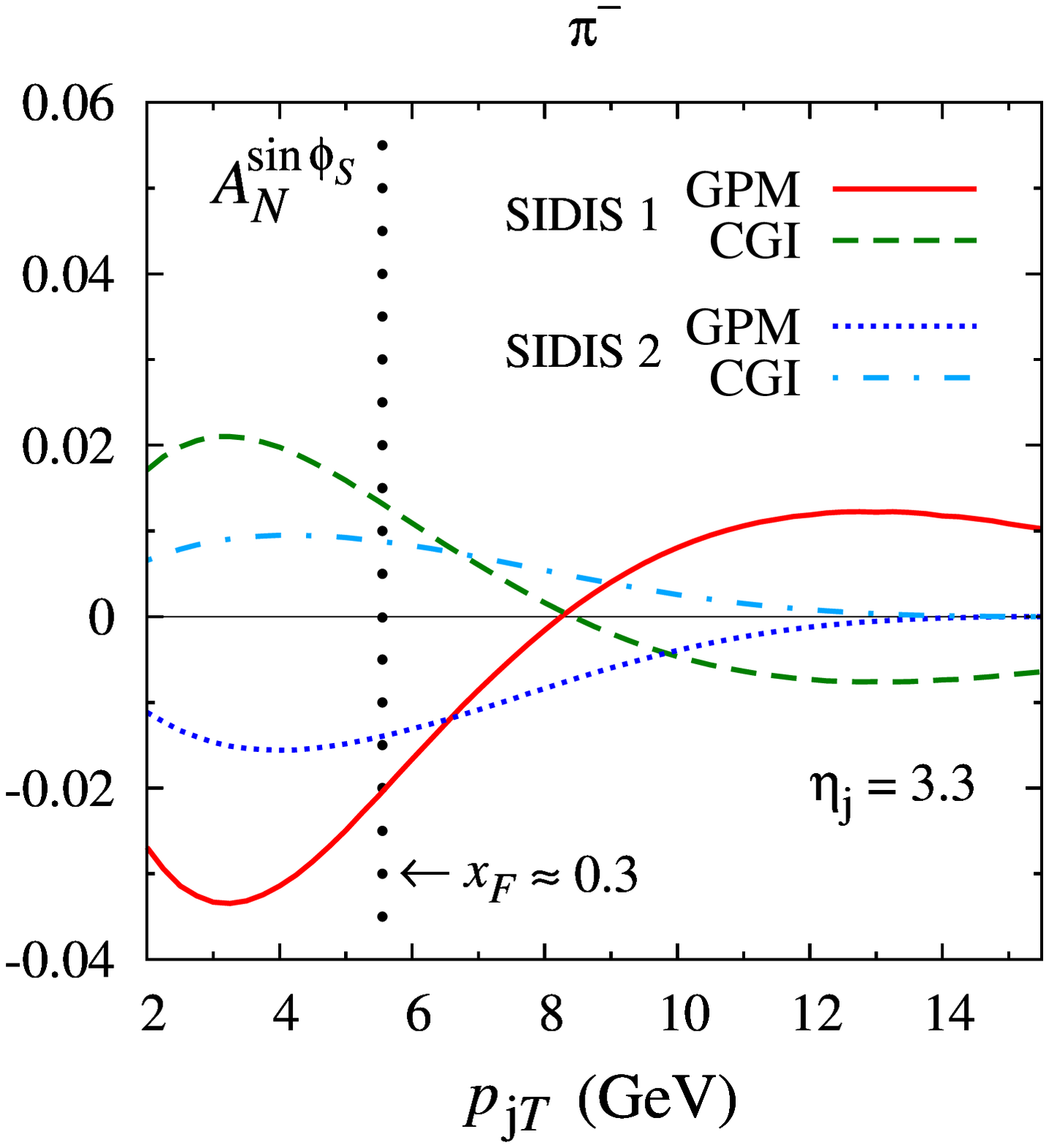}
\includegraphics[width=5.5cm]{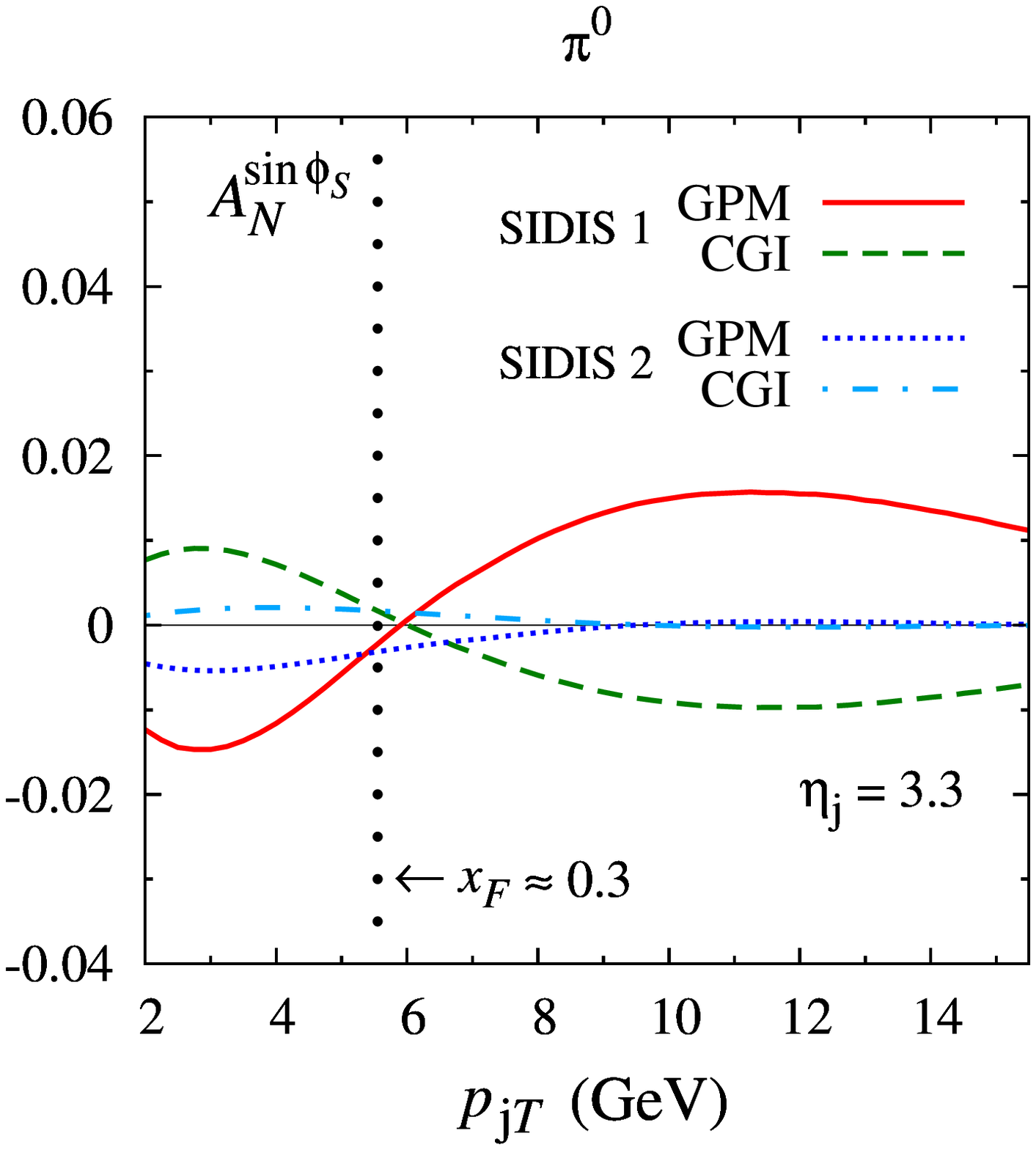}
\includegraphics[width=5.5cm]{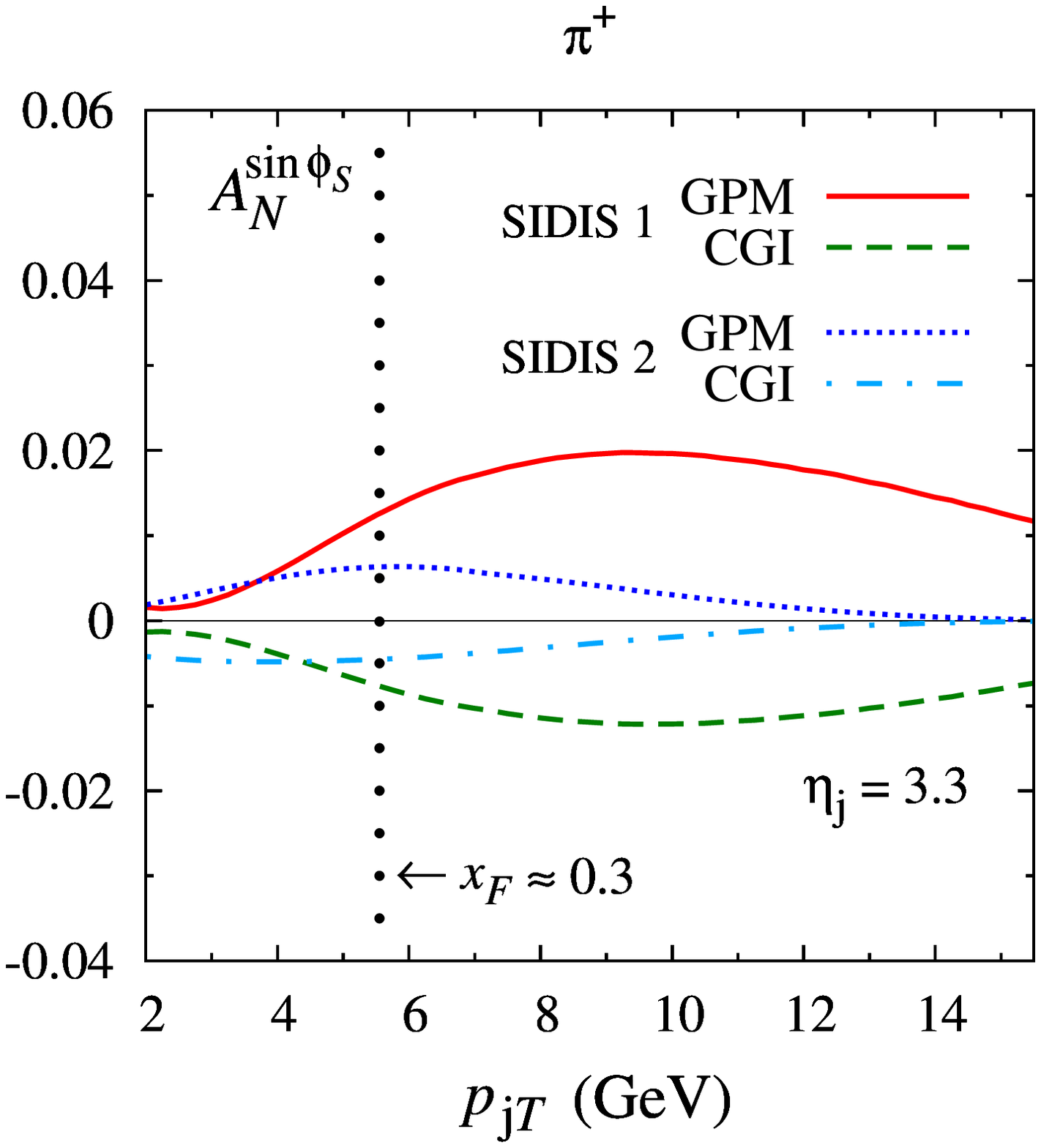}
\caption{Quark contribution to the Sivers asymmetry
$A_N^{\sin\phi_{S}}$ in the GPM and CGI-GPM frameworks for the process $p^\uparrow \, p\to {\rm jet}\, \pi\,X$, at the energy $\sqrt{s}= 500$ GeV and fixed value 
of the jet rapidity $\eta_{\rm j} = 3.3$, as a function of $p_{{\rm j} T}$.
Estimates are obtained by adopting the parametrization sets SIDIS~1 and 
SIDIS~2.}
\label{asy-an-siv-par500}
\end{figure*}

\begin{figure*}[t]
\centering  
\includegraphics[width=5.5cm]{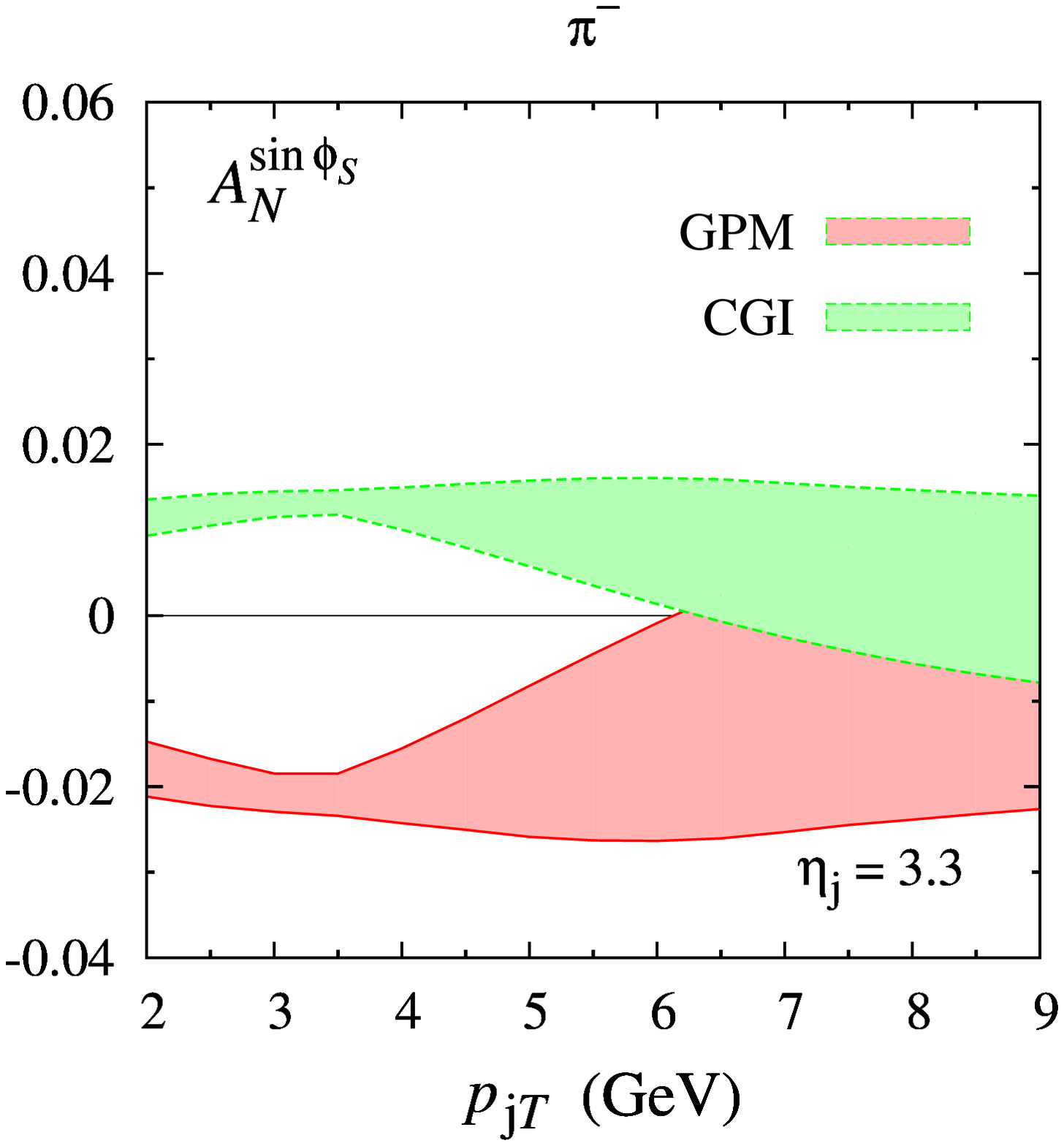}
\includegraphics[width=5.5cm]{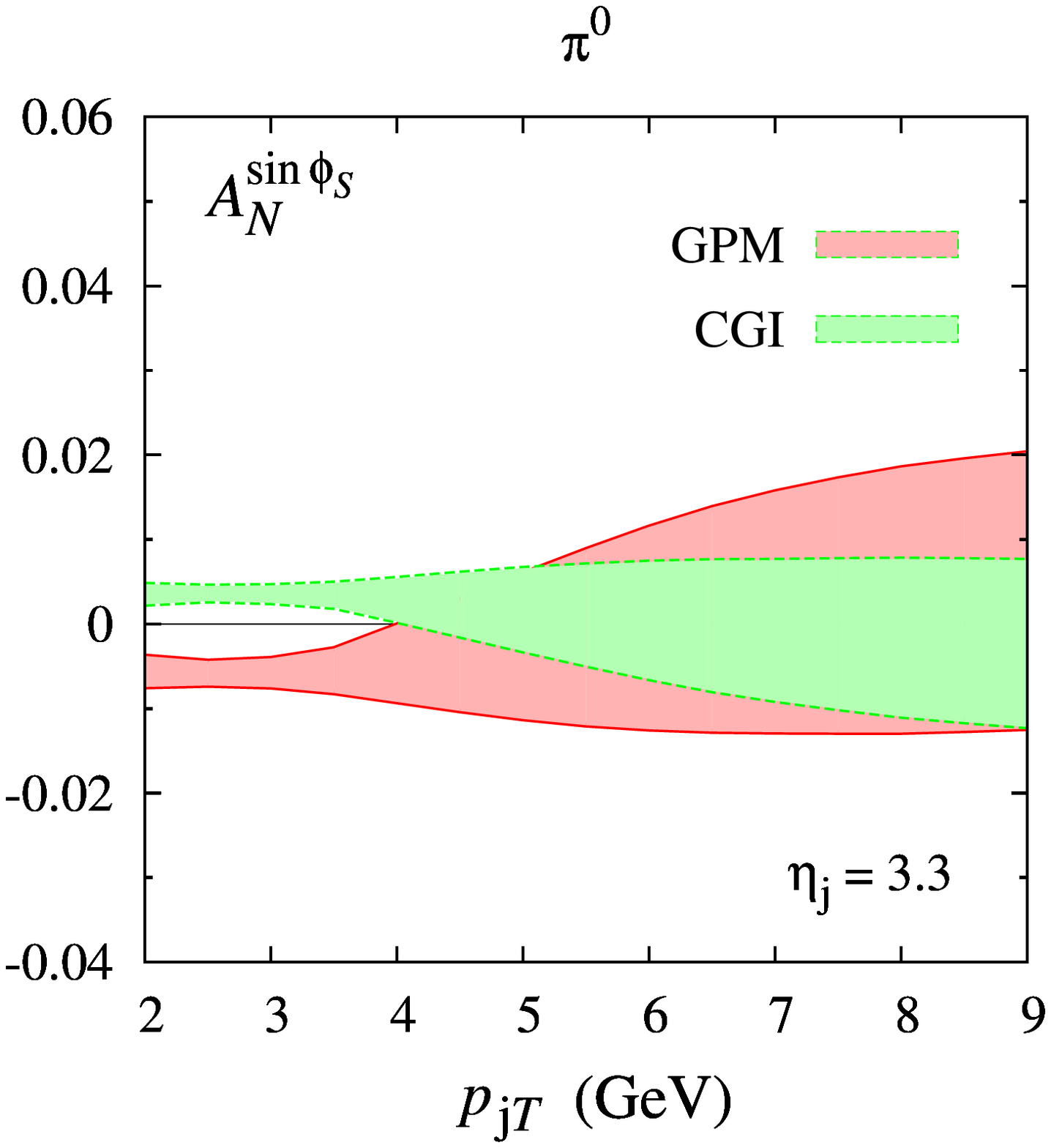}
\includegraphics[width=5.5cm]{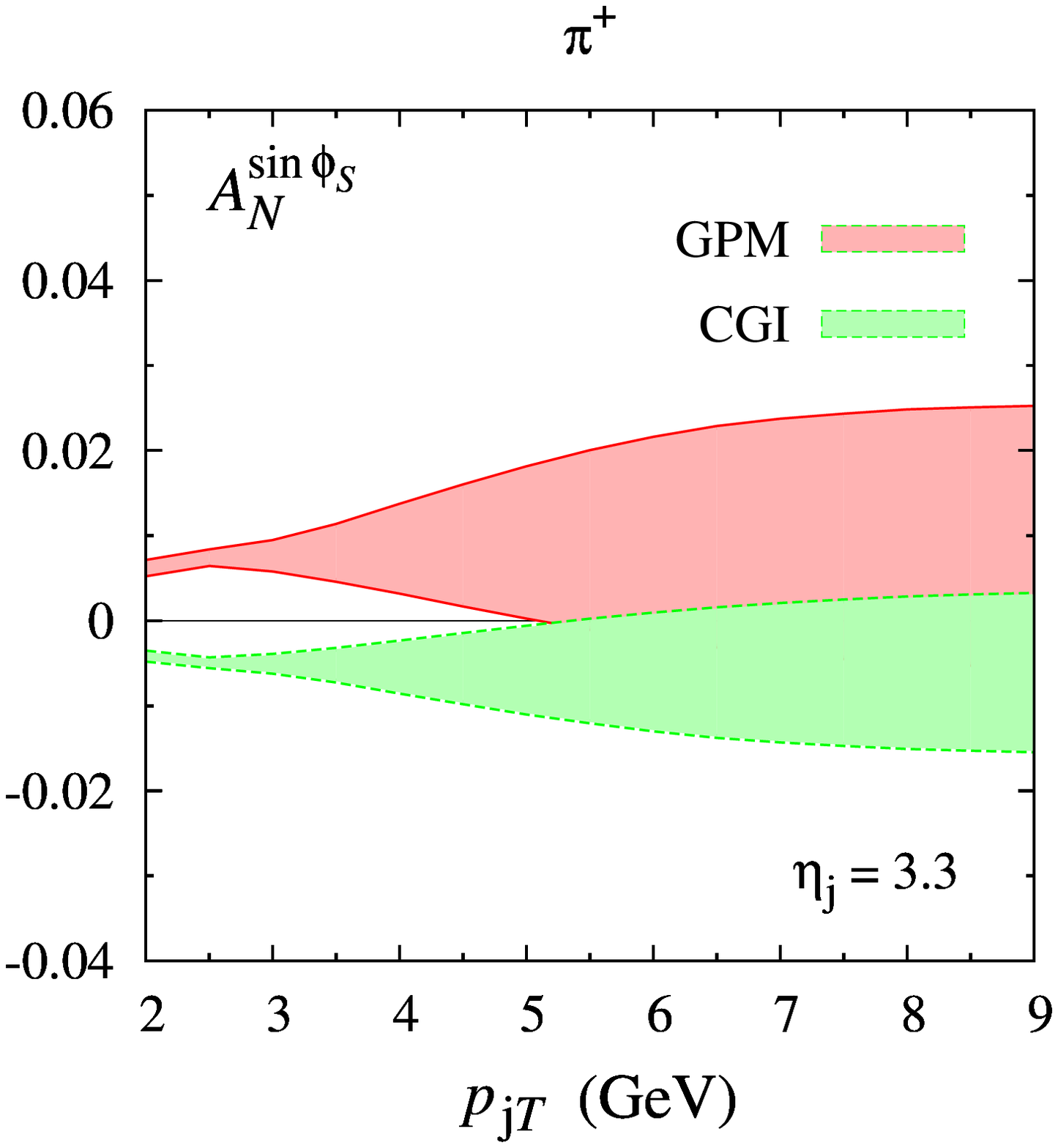}
\caption{Scan bands describing the uncertainty of the
quark contribution to the Sivers asymmetry $A_N^{\sin\phi_{S}}$ in the GPM and 
CGI-GPM frameworks for the process $p^\uparrow \, p\to {\rm jet}\, \pi\,X$,
 at $\sqrt{s}= 500$ GeV and fixed value of the jet rapidity 
$\eta_{\rm j} = 3.3$, as a function of $p_{{\rm j} T}$.}
\label{asy_siv_scan_500}     
\end{figure*}

In Fig.~\ref{fig:coll-scan} we show our resulting scan bands for the
Collins azimuthal asymmetry $A_N^{\sin(\phi_{S}- \phi_\pi^H)}$ for neutral and
charged pions at the RHIC center-of-mass energy $\sqrt{s}=500$ GeV and fixed jet pseudorapidity $\eta_{\rm j}=3.3$, as a function of the jet transverse momentum. This envelope provides an estimate of the uncertainty in the asymmetry calculation which increases as $p_{{\rm j} T}$ and $x_F$ (at fixed $\eta_{\rm j}$) grow. This information integrates the indication obtained comparing the results of the SIDIS~1 and SIDIS~2 sets in Fig.~\ref{asy-an-coll-par200}.

We point out that these asymmetries are currently under active investigation by the STAR Collaboration \cite{Poljak:2011vu,Fatemi:2012ry} in the central rapidity region as well, where they turn out to be much smaller. Finally, we cannot 
provide similar estimates for the azimuthal moment $A_N^{\sin(\phi_{S}- 2 \phi_\pi^H)}$, because the underlying TMD gluon distribution and fragmentation functions
are, at the moment, completely unknown.

\subsection{The Sivers Asymmetries}
\label{sec:siv}

Similarly to Eqs.~(\ref{eq:Coll1}) and (\ref{eq:Coll2}), the expression for the
a\-zi\-mu\-thal moment $A_N^{\sin\phi_S}$ is schematically given by  
\begin{equation}
A_N^ {\sin\phi_S}\sim  f_{1 T}^{\perp}(x_a,\bm{k}_{\perp a}^2) \otimes
 f_1(x_b,\bm{k}_{\perp b}^2)  \otimes D_1(z,\bm{k}_{\perp \pi}^2)\,,
\label{eq:Siv}
\end{equation}
where the Sivers function $f_{1 T}^{\perp}$ for an unpolarized parton $a$ inside the transversely polarized proton is convoluted with the unpolarized distribution $f_1$ for parton $b$ and the fragmentation function $D_1$ for parton $c$.  

The Sivers asymmetry for charged and neutral pions is presented in Fig.~\ref{asy-an-siv-par200} at the center-of-mass energy $\sqrt{s}=200$ GeV and at forward rapidity $\eta_{\rm j}=3.3$, as a function of $p_{{\rm j} T}$. The quark and gluon contributions are depicted  separately, although in principle it is not possible to disentangle them. However it should be possible to identify specific kinematic regions in which only one of them dominates. The almost unknown gluon Sivers function is assumed to be positive and to saturate an updated version of the bound in Ref.~\cite{Anselmino:2006yq}. Such bound has been derived from the analysis of PHENIX data on transverse single spin asymmetries for the process $p^{\uparrow}p\to \pi^0\,X$, with the neutral pion produced at central rapidities. The quark Sivers function is estimated by adopting the SIDIS~1 and SIDIS~2 parameterizations. As  for the case of the Collins asymmetry, predictions are comparable only in the $p_{{\rm j} T}$ region delimited by the dotted vertical line, where our parameterizations are constrained by SIDIS data.
The measurement of $A_N^{\sin\phi_{S}}$ at large $p_{{\rm j} T}$,
where the role of the gluon Sivers function is negligible, could help to 
discriminate between the two parameterizations and constrain the 
behaviour of the $u$, $d$ quark Sivers functions at large $x$ .

The present analysis can be easily extended to the transverse single spin asymmetry for inclusive jet production in $p^\uparrow  p\to {\rm jet}\,X$, by simply integrating the results for  $p^\uparrow p\to {\rm jet}\,\pi\,X$  over the pion phase space. In this case, in the angular structure of the asymmetry in Eq.~(\ref{num-asy-gen}) only the $\sin\phi_S$ modulation will be present, because all
the mechanisms related to the fragmentation process cannot play a role. Our 
predictions for $A_N^{\sin\phi_{S}}$ turn out to be very similar to 
the ones for jet-neutral pion production, presented in the central panel 
of Fig.~\ref{asy-an-siv-par200}.

\section{The Sivers Asymmetry in the Color Gauge Invariant GPM}

In the GPM framework adopted so far, TMD distribution and 
fragmentation functions are taken to be universal. This is generally 
believed to be the case for the Collins function, at least for the processes in 
which QCD factorization has been estabilished. On the other hand, several 
naively time-reversal odd TMD distributions, like for example the Sivers
function, can depend on initial (ISIs) or/and final (FSIs) state interactions 
between the struck parton and the soft remnants in the process. Such 
interactions depend on the particular reaction under study and can render the
TMD distribution non-universal. A fundamental example (still lacking 
experimental evidence) is provided by the effects of ISIs in SIDIS and FSIs 
in the DY processes, which lead to two different quark Sivers functions with an 
opposite relative sign. These effects are taken into account in the 
color gauge invariant (CGI) GPM approach~\cite{Gamberg:2010tj,D'Alesio:2011mc}. 
For the process $p^{\uparrow}p\to {\rm jet}\,\pi\,X$, the quark Sivers function 
has in general a more involved color structure as compared to the SIDIS and DY 
cases, since both ISIs and FSIs can in principle contribute \cite{D'Alesio:2011mc}. However, the situation becomes simpler at forward rapidities, where
only the $qg\to qg$ channel dominates. As a consequence, our predictions for 
the Sivers asymmetries, calculated with and without ISIs and FSIs, are 
comparable in size but have opposite signs. 

Our results for $A_N^{\sin\phi_S}$ are  depicted in Fig.~\ref{asy-an-siv-par500} at the RHIC energy $\sqrt{s} = 500$ GeV using the two available sets SIDIS~1 and SIDIS~2 for the quark Sivers function. It is clear from the picture that the 
 measurement of a sizable asymmetry would validate one of the two approaches and test the process dependence of the Sivers function.
These conclusions are confirmed by Fig.~\ref{asy_siv_scan_500}, where
scan bands for the asymmetries have been obtained following the same procedure
described for the Collins effect in Section~\ref{sec:coll} and in Refs.~\cite{Anselmino:2012rq,Anselmino:2013rya}.

\begin{figure}
\centering
\includegraphics[width=7cm]{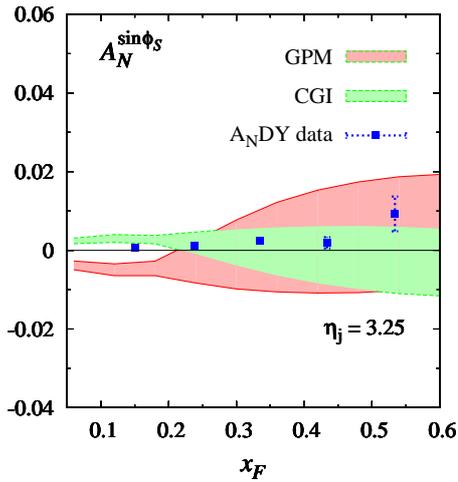}
\caption{Scan bands describing the uncertainty of the quark 
contribution to the Sivers asymmetry $A_N^{\sin\phi_{S}}$  in the GPM and 
CGI-GPM frameworks, for the process $p^\uparrow \, p\to {\rm jet}\,X$ 
at the energy $\sqrt{s}= 500$ GeV and at fixed value of the jet rapidity $\eta_{\rm j}=3.25$, as a function of $x_F$.}
\label{fig:andy}
\end{figure}

Finally, we have studied $A_N^{\sin\phi_{S}}$  for inclusive jet production 
\cite{D'Alesio:2010am,D'Alesio:2011mc}. Similarly to our predictions in the GPM
approach, also in the CGI GPM framework $A_N^{\sin\phi_{S}}$  turns out to be 
very similar to the one for jet-neutral pion production, shown in the central 
panel of Fig.~\ref{asy-an-siv-par500}. According to the data reported by the 
A$_N$DY Collaboration at RHIC, presented in Fig.~\ref{fig:andy}, 
the Sivers asymmetry for $p^\uparrow p\to {\rm jet}\,X$ is small and 
positive \cite{Nogach:2012sh,Bland:2013pkt}. In the same figure we show
also the scan bands for the quark Sivers asymmetries evaluated in the GPM and 
CGI-GPM frameworks. The GPM predictions agree with the data only for 
$x_F \ge 0.3$. This suggest the need for further studies along these lines, 
aiming to confirm or disprove the validity of our TMD factorization
assumption and to investigate the universality properties of the Sivers 
function.  

\section{Concluding Remarks}

In the framework of the generalized parton model, we have discussed the 
phenomenological relevance and usefulness of the process 
$p^{\uparrow} p\to {\rm jet}\,\pi\,X$ for the study  of TMD parton distributions and fragmentation functions. In particular, we have shown how our proposed measurents can shed light on the large $x$ behaviour of the TMD quark transversity distributions and of the quark Sivers functions, complementing information coming from other reactions like SIDIS, Drell-Yan and $e^+e^-$ annihilations. We have also presented an additional phenomenological study of the process dependence of the Sivers function for quarks. Comparison with experiments will allow us to test our hypothesis on the validity of TMD factorization, and to assess the role and size of possibile factorization-breaking terms.

\section*{Acknowledgements}
U.D.\ and F.M.\ acknowledge financial support from the European Community under the FP7 “Capacities - Research Infrastructures” programme (HadronPhysics3, Grant Agreement 283286). C.P.\ is supported by the European Community under the Ideas program QWORK (contract 320389). U.D.\ is grateful to the Department of Theoretical Physics II of the Universidad Complutense of Madrid for the kind hospitality extended to him during the completion of this work. \\


\begin{thebibliography}{}

\bibitem{D'Alesio:2007jt}
U.~D'Alesio, F.~Murgia,
Prog. Part. Nucl. Phys. {\textbf{61}}, 394 (2008), \texttt{0712.4328}

\bibitem{Barone:2010zz}
V.~Barone, F.~Bradamante, A.~Martin,
Prog. Part. Nucl. Phys. {\textbf 65}, 267 (2010), \texttt{1011.0909}

\bibitem{Sivers:1989cc}
D.W~Sivers, Phys.\ Rev.\ \textbf{D41}, 83 (1990); {\it{ibidem}} \textbf{D43}, {261} (1991)

\bibitem{Boer:1997nt}
D.~Boer, P.J.~Mulders, Phys.\ Rev.\ \textbf{D57}, 5780 (1998), 
\texttt{hep-ph/9711485}

\bibitem{Collins:1992kk}
J.C.~Collins, Nucl.\ Phys.\ \textbf{B396}, 161 (1993), \texttt{hep-ph/9208213}

\bibitem{Mulders:2000sh} 
  P.~J.~Mulders, J.~Rodrigues,
  Phys.\ Rev.\ \textbf{D63}, 094021 (2001), \texttt{hep-ph/0009343}

\bibitem{D'Alesio:2010am}
U.\ D'Alesio, F.\ Murgia,  C.\ Pisano,  Phys.\ Rev.\ \textbf{D83}, 
034021 (2011), \texttt{1011.2692}

\bibitem{D'Alesio:2011mc}
U.\ D'Alesio, L.\ Gamberg, Z.B.\ Kang, F.\ Murgia, C.\ Pisano,
Phys. Lett.  \textbf{B704}, 637 (2011), \texttt{1108.0827}

\bibitem{D'Alesio:2013jka}
  U.\ D'Alesio, F.\ Murgia, C.\ Pisano,
  Phys.\ Part.\ Nucl.\  \textbf{45}, no.\ 4, 676 (2014), \texttt{1307.4880}

\bibitem{Poljak:2011vu}
N.~Poljak (STAR Collaboration),
Nuovo Cim. \textbf{C35}, 193 (2012), \texttt{1111.0755}

\bibitem{Fatemi:2012ry}
R.~Fatemi (STAR Collaboration),
 AIP Conf. Proc. \textbf{1441}, 233 (2012), \texttt{1206.3861}

\bibitem{Yuan:2007nd} 
  F.~Yuan,
  Phys.\ Rev.\ Lett.\  \textbf{100}, 032003 (2008), \texttt{0709.3272}

\bibitem{Anselmino:2005ea} 
  M.~Anselmino, M.~Boglione, U.~D'Alesio, A.~Kotzinian, F.~Murgia, A.~Prokudin,
  Phys.\ Rev.\ \textbf{D72}, 094007 (2005), \texttt{hep-ph/0507181}

\bibitem{Anselmino:2007fs} 
  M.~Anselmino, M.~Boglione, U.~D'Alesio, A.~Kotzinian, F.~Murgia, A.~Prokudin, C.~Turk, Phys.\ Rev.\ \textbf{D75}, 054032 (2007), \texttt{hep-ph/0701006}

\bibitem{Anselmino:2008sga} 
  M.~Anselmino, M.~Boglione, U.~D'Alesio, A.~Kotzinian, S.~Melis, F.~Murgia, A.~Prokudin, C.~Turk, Eur.\ Phys.\ J.\  \textbf{A39}, 89 (2009), \texttt{0805.2677}

\bibitem{Anselmino:2008jk} 
  M.~Anselmino, M.~Boglione, U.~D'Alesio, A.~Kotzinian, F.~Murgia, A.~Prokudin, S.~Melis,
  Nucl.\ Phys.\ Proc.\ Suppl.\  \textbf{191}, 98 (2009), \texttt{0812.4366}

\bibitem{Anselmino:2013vqa}
M.~Anselmino, M.~Boglione, U.~D'Alesio, S.~Melis, F.~Murgia, A.~Prokudin,
Phys.\ Rev.\ \textbf{D87}, 094019 (2013), \texttt{1303.3822}

\bibitem{Anselmino:2012rq}
M.~Anselmino, M.~Boglione, U.~D'Alesio, E.~Leader, S.~Melis, F.~Murgia, 
A.~Prokudin, Phys. Rev. {\bf D86}, 074032 (2012), \texttt{1207.6529}


\bibitem{Anselmino:2006yq}
M.~Anselmino, U.~D'Alesio, S.~Melis, F.~Murgia,
Phys. Rev. {\bf D74}, 094011 (2006), \texttt{hep-ph/0608211}

\bibitem{Gamberg:2010tj}
L.~Gamberg, Z.B. Kang, Phys. Lett.\ \textbf{B696}, 109 (2011), 
\texttt{1009.1936}

\bibitem{Anselmino:2013rya}
M.~Anselmino, M.~Boglione, U.~D'Alesio, S.~Melis, F.~Murgia, A.~Prokudin,
Phys.\ Rev.\ \textbf{D88}, 054023 (2013), \texttt{1304.7691}

\bibitem{Nogach:2012sh}
L.~Nogach (A$_N$DY Collaboration), \texttt{1212.3437}

\bibitem{Bland:2013pkt}
L.~Bland {\em et~al.} (A$_N$DY Collaboration), \texttt{1304.1454}

\end{thebibliography}
\end{document}